%% file: 0-main.tex
\documentclass[trackchanges,twocolumn]{aastex631}
\usepackage[utf8]{inputenc}
\usepackage[english]{babel}
\usepackage{CJKutf8}
\usepackage{graphicx, float, multirow}  
\usepackage{enumerate, enumitem, hyperref, url, xcolor}     
\usepackage{amssymb, amsmath, wasysym, chemformula, chemfig}         
\let\tablenum\relax
\usepackage[group-digits = none]{siunitx}
\usepackage{booktabs}

\newcommand{\dff}{\textrm{d}} 

\newcommand{\dfp}{\partial}
\newcommand{\Omgstar}{\Omega^*}

\shorttitle{Planetary Rotation and Seasonality of Clouds}
\shortauthors{D. A. Williams et al.}

\begin{document}
\begin{CJK*}{UTF8}{gbsn}

\title{Clouds and Seasonality on Terrestrial Planets with Varying Rotation Rates}

\author[0000-0002-5840-2411]{Daniel A. Williams}
\affiliation{Department of Mathematics and Statistics, Faculty of Environment, Science and Economy, University of Exeter, UK}

\author[0000-0002-1592-7832]{Xuan Ji(纪璇)}

\affiliation{Department of the Geophysical Sciences, University of Chicago, IL, USA}

\author[0000-0002-6417-9316]{Paul Corlies}
\affiliation{Department of Earth, Atmospheric, and Planetary Sciences, Massachusetts Institute of Technology, Cambridge, MA, USA}

\author[0000-0001-9925-1050]{Juan M. Lora}
\affiliation{Department of Earth and Planetary Sciences, Yale University, New Haven, CT, USA}

\correspondingauthor{Daniel A. Williams, Juan M. Lora}
\email{dw569@exeter.ac.uk, juan.lora@yale.edu}

\begin{abstract}
Using an idealised climate model incorporating seasonal forcing, we investigate the impact of rotation rate on the abundance of clouds on an Earth-like aquaplanet, and the resulting impacts upon albedo and seasonality. We show that the cloud distribution varies significantly with season, depending strongly on the rotation rate, and is well explained by the large-scale circulation and atmospheric state. Planetary albedo displays non-monotonic behaviour with rotation rate, peaking at around $1/2\Omega_E$. Clouds reduce the surface temperature and total precipitation relative to simulations without clouds at all rotation rates, and reduce the dependence of total precipitation on rotation rate, causing non-monotonic behaviour and a local maximum around $1/8\Omega_E$; these effects are related to the impacts of clouds on the net atmospheric and surface radiative energy budgets. Clouds also affect the seasonality. The influence of clouds on the extent of the winter Hadley cell and the intertropical convergence zone is relatively minor at slow rotation rates ($< 1/8\Omega_E$), but becomes more pronounced at intermediate rotation rates, where clouds decrease their maximum latitudes. The timing of seasonal transitions varies with rotation rate, and the addition of clouds reduces the seasonal phase lag.
\end{abstract}


\input{1-introduction}
\input{2-methods}
\input{3-results}
\input{4-discussion}

\newpage
We give thanks to the organisers of the Rossbypalooza 2022 Summer School at the University of Chicago, at which the investigation that led to this manuscript started. DAW was supported by an STFC studentship  (ST/V506667/1); this work was partially supported by NASA grant No. 80NSSC21K1718, which is part of the Habitable Worlds program. For the purpose of open access, the author has applied a Creative Commons Attribution (CC BY) licence to any Author Accepted Manuscript version arising. The research data supporting this publication are openly available at \href{https://doi.org/10.5281/zenodo.8075444}{\texttt{doi://10.5281/zenodo.8075444}}.

We thank T.~Komacek for providing the output data from ExoCAM simulations for comparison. We also extend thanks to a number of people who offered constructive comments and ideas during the undertaking of this work, in alphabetical order: D.~Abbot, M.~Battalio, B.~Fan, E.~Kite, N.~Lewis, D.~Sergeev, W.~Seviour, S.~Thomson, G.~Vallis, and A.~Warren. This work was completed in part with resources provided by the University of Chicago Research Computing Center. We finally would like to thank two reviewers for their help in clarifying and improving the paper.

\software{Isca \citep{vallis2018-Isca},  
          xarray \citep{hoyer2022-Xarray}, 
          cartopy \citep{elson2022-SciTools}
          }

\bibliography{rossbypalooza}{}
\bibliographystyle{aasjournal}

\end{CJK*}
\end{document}

%% file: 1-introduction.tex
\section{Introduction}\label{sec:intro}

The past 30 years have brought the discovery of roughly 5000 planets beyond our solar system, most detected by the NASA Kepler \citep{borucki2011-Characteristics} and TESS \citep{dragomir2019-TESS, cacciapuoti2022-TESS} missions. Amongst the catalogue of confirmed exoplanets, the majority are larger and more massive than Earth \citep{nasaexoplanetscienceinstitute2020-Planetary}; biases in  detection methods make large, rapidly-orbiting planets easier to detect \citep{burrows2014-Highlights}. Nevertheless, increasing numbers of smaller terrestrial planets have been detected \citep{bryson2020-Occurrence}. The TRAPPIST-1 system \citep{gillon2017-Seven} is perhaps the most publicised due to it hosting a large number of planets, but recently more exoplanets very similar to Earth in size and equilibrium temperature have been detected \citep{gilbert2020-First}. With the development of improved analytical methods and the launch of the latest generation of telescopes, it is now possible to characterise the composition of exoplanetary atmospheres \citep[e.g.,][]{alderson2022-Early}, greatly enhancing the possibility of using climate models to understand these exotic worlds.

A key aspect of planetary atmospheres is the presence or absence of clouds. The existence of clouds relies on the presence of a gaseous chemical species in a planet's atmosphere that can condense into cloud particles. Clouds on Earth are a familiar concept, but it is not the only planet where they are found. Beyond Earth, clouds of varying nature have been observed on other terrestrial bodies, both within our solar system and beyond. Our nearest neighbours Venus and Mars both have observable clouds: Venus's take the form of a thick global layer of sulphuric acid \citep{krasnopolsky1981-Chemical}; on Mars, tenuous water ice and carbon dioxide ice clouds have been observed \citep[e.g.,][and references therein]{curran1973-Mars, belliii1996-Detection, haberle2017-Atmosphere}. Saturn's moon Titan hosts a great variety of clouds, with a varying seasonal distribution, morphology, and composition \citep[e.g.,][]{griffith2006-Titan, dekok2014-HCN, turtle2018-Titan}.

Using visible and near-infrared spectra from telescopes such as the Very Large Telescope, the Hubble Space Telescope and the James Webb Space Telescope, we have been able to infer the presence of clouds on exoplanets \citep{kreidberg2014-Clouds, sing2016-Continuum, samland2017-Spectral, barstow2021-Curse}. For instance, \citet{alderson2022-Early} inferred the existence of clouds in the atmosphere of WASP-39b through the continuum transit depth observed with JWST. The sheer variety of exoplanets thus far discovered has also indicated the presence of exotic condensing species, such as metal oxides and silicates in the atmospheres of very hot planets \citep[e.g.,][]{helling2019-Exoplanet}.

Clouds have a significant impact on the climate of a planet, both through their influence on the transport of condensable species in the atmosphere and on the radiation budget of the climate system \citep{trenberth2012-Tracking}. Clouds act as a source and sink of moisture, which, via latent heating associated with phase transitions, modify the temperature and stability of their atmospheric environment. Their optical properties as effective scattering media have an impact on both incoming short-wave radiation (affecting planetary albedo) and long-wave radiation emitted by the planet; these cloud radiative feedbacks are broadly understood \citep{webster1994-Role, pierrehumbert2010-Principles} and can be reasonably represented in models \citep{rose2021-Climate}, though clouds also represent one of the larger uncertainties in state-of-the-art models of Earth's climate \citep{ceppi2017-Cloud}. Cloud formation is influenced by a number of factors, chiefly the temperature and specific humidity of an air parcel. These quantities are themselves influenced by processes such as convection and advection, such that large-scale atmospheric dynamics are a significant driver of global cloud distribution.

Despite their importance to the energy budget of the climate system, clouds are often neglected in planetary climate models due to difficulties in modelling them accurately. Clouds present a range of complex microphysical processes that are not fully understood: consider the formation of different ice crystals \citep{storelvmo2015-WegenerBergeronFindeisen}, the physics of mixed-phase clouds \citep{korolev2003-Supersaturation}, or the subsequent formation and growth of raindrops \citep{bergeron1935-Physics}. In addition, the processes and dynamics within clouds occur at small scales, beyond the spatial resolution of general circulation models (GCMs). These sub-grid processes must be parameterised in order to be represented in such models, and there is great variation between cloud schemes of different GCMs. Further, cloud schemes add an additional computational cost to any model. As a result, clouds are often ignored and excluded, even in simplified models of the Earth \citep{thomson2019-Hierarchical}. Nonetheless, given the clear evidence of clouds on other bodies within our own solar system and increasing observations and inferences of exoplanetary clouds, as well as their critical importance to Earth's climate system, it is pertinent to attempt to better account for clouds' impacts in models, especially given their potential importance to observable quantities such as albedo and emission or transmission spectra.

Separately, a number of previous studies have investigated how the atmospheric circulation and climate of a terrestrial planet may be influenced by different atmospheric and planetary characteristics. Planetary rotation rate is well understood as a key determiner of atmospheric dynamics, affecting the width and magnitude of the Hadley circulation \citep{kaspi2015-ATMOSPHERIC, guendelman2018-Axisymmetric, singh2019-Limits, hill2022-Theory}, latitudinal distribution of extratropical eddies \citep{eady1949-Long, taylor1980-Roles}, related extratropical jets \citep{williams1978-Planetary, cho1996-Emergence, chemke2015-Latitudinal}, heat transport\citep{liu2017-effect, cox2021_radiative}, radiative cooling \citep{zhang2023_inhomogeneity}, and general climate and habitability \citep{yang2014-STRONG, haqq-misra2018-Demarcating, komacek2019-Atmospheric, jansen2019-Climates, guzewich2020, cox2021_radiative, he2022-Climate}. The role of rotation rate has often been studied using idealised GCMs, which strip out factors such as topography in order to understand the fundamental physics that drive the dynamics and climate \citep{schneider2006-General, ogorman2008-Hydrological, thomson2019-Effects}. More broadly, \citet{kaspi2015-ATMOSPHERIC} used a number of parameter sweeps to investigate the effects of rotation rate, mean insolation, atmospheric mass, atmospheric optical density, planetary density and radius on the resulting atmospheric circulation in an idealised, Earth-like GCM. Their simulations were based on a clear-sky, equinoctinal aquaplanet with a grey radiation scheme, constructed using the Flexible Modeling System developed at GFDL \citep{held1994-Proposal, anderson2004-New, frierson2006-GrayRadiation}. They found that rotation rate is a key driver of atmospheric dynamics, describing two principal regimes: fast rotators, characterised by a weaker Hadley circulation and development of extratropical jets; and slow rotators, with strong Hadley circulation and a small meridional temperature gradient. \citet{ogorman2008-Hydrological} used a similar grey radiation model to study the connection between the global mean temperature and the hydrological cycle, finding that, as the climate warms, the global-mean precipitation eventually reaches an asymptotic value. Whilst these models included moisture and a representation of the hydrological cycle, all excluded a cloud scheme. 

Various past efforts have used GCM simulations with models of higher complexity to interrogate clouds. For example, \citet{parmentier2016-TRANSITIONS} compared cloud presence in exoplanet transit data of hot Jupiters to models building on the thermal structure from GCMs. They also suggested that cloud presence and composition may be deduced from asymmetries in the light curves of transiting cool exoplanets, providing a potential basis for diagnosing clouds from observational data. \citet{komacek2019-Atmospheric} built on the previous work by \citet{kaspi2015-ATMOSPHERIC}, making use of the \textsc{ExoCAM} model \citep{wolf2022-ExoCAM}, to perform a range of GCM simulations including sea ice, a correlated-k radiation scheme, and a cloud scheme \citep{rasch1998-Comparison}. Their findings were qualitatively consistent with the results of \citet{kaspi2015-ATMOSPHERIC}, but they observed a larger equator-to-pole temperature gradient. This may be attributed to their use of a more complex radiative scheme and addition of a cloud scheme; they also noted that cloud particle size appears to be an important but uncertain parameter. While they presented an interesting transition between low and high day-side cloud coverage for synchronously rotating planets with varying rotation periods, the impact of clouds on atmospheric dynamics and climate for asynchronous planets was not explicitly addressed. \citet{yang2013-STABILIZING} and \citet{yang2014-STRONG} also investigated clouds over a range of rotation rates, including tidally-locked systems. They suggested that the planetary albedo increases to a maximum in the synchronous case, where a substantial cloud cap forms at the sub-stellar point. Finally, \citet{guzewich2020} showed that variations of cloud properties with rotation rate can alter the observable signatures of otherwise Earth-like planets in reflected light and thermal emission spectra.

Most of the previous studies generally considered planets without a seasonal cycle, hence providing a hemispherically symmetric circulation. Whilst simulating a planet with no obliquity ($\varepsilon=0$) has advantages, we know from our own solar system that planets rarely conform to this simplification, as Earth, Mars and Titan all have large effective obliquities of approximately $\varepsilon=25^\circ$. This results in seasonally-varying radiative forcing that is a key driver of dynamics that are absent in aseasonal climates \citep{guendelman2018-Axisymmetric, guendelman2019-Atmospheric, ohno2019_atmospheres, singh2019-Limits, hill2022-Theory}, including the Earth's monsoon systems \citep{bordoni2008-Monsoons, geen2018-Regime}; in this context, for example, the migration of the Intertropical Convergence Zone (ITCZ) has been shown to vary non-monotonically with rotation rate \citep{faulk2017-Effects,geen2018-Regime}. Other studies have explored how obliquity influences the climatological temperature, precipitation and habitability \citep{ferreira2014-Climate, kang2019-Mechanisms, kodama2022-Climate, linsenmeier2015-Climate, lobo2020-Atmospheric, he2022-Climate}. Furthermore, \citet{lobo2022-Role} analysed the relationship between extratropical storminess and longwave radiation, and \citet{hadas2023-Role} took cloud albedo into consideration, linking Earth's large-scale circulation to planetary albedo, which indicates the importance of using a more realistic radiative scheme to study the effect of rotation rate.

This work builds on the above studies, especially those by \citet{kaspi2015-ATMOSPHERIC} and \citet{komacek2019-Atmospheric}, by interrogating the effects of clouds on varying planetary climates with idealised GCM simulations including a seasonal cycle. By covering a large range of rotation rates, we aim to include a swathe of dynamical regimes with a likely impact on cloud distributions, enabling us to investigate the response of seasonal cloud behaviour to rotation rate and the feedback of these clouds on the seasonal climate. We do not consider tidally locked systems where rotation and orbital rate are synchronous; tidally locked planets display a distinct dynamical regime \citep{joshi1997-Simulations, noda2017-Circulation, haqq-misra2018-Demarcating, pierrehumbert2019-Atmospheric, wordsworth2022-Atmospheres} where dynamics are not longitudinally invariant \citep{merlis2010-Atmospheric, sergeev2020-Atmospheric, hammond2021-Rotational}, and the presence and influence of clouds on tidally locked planets have already been investigated extensively \citep{yang2013-STABILIZING, yang2019-Simulations, helling2021-Clouds, sergeev2022-TRAPPIST1}.

Hereon, we describe the experimental setup and procedure in Section 2. Following this we present our results, starting with the effect of rotation rate on the dynamics and large-scale distribution of clouds, including their impacts on planetary albedo (Section 3.1). We follow this with results investigating the consequences of clouds upon climate and seasonality (Section 3.2), and discuss comparisons with previous work and provide a future outlook in Section 4.

%% file: 2-methods.tex
\section{Methods}\label{sec:methods}

\subsection{Idealised GCM Setup}
In order to explore the effect of planetary rotation rate on the dynamics of a simplified cloudy aquaplanet, we make use of the flexible modelling framework Isca \citep{vallis2018-Isca}. Isca is not a single model, but a framework based on the GFDL Flexible Modeling System for constructing a range of idealised GCMs covering a wide hierarchy of complexity \citep{thomson2019-Hierarchical}, allowing the effects of particular processes to be studied in isolation. Complexity may range from a very simple Newtonian thermal-relaxation model as described by \citet{held1994-Proposal} to models that introduce additional complexity such as moist physics, topography and differing radiation schemes. 

The initial base for our setup in Isca can be derived from the moderately complex Earth-like model described in \citet{thomson2019-Hierarchical}. We use a 360-day calendar with all planetary parameters identical to Earth except rotation rate $\Omega$. We also use a simplified orbit, including the Earth's obliquity but neglecting eccentricity such that seasonal forcing is symmetrical across the northern/southern hemispheres.

Isca uses a spectral dynamical core in spherical coordinates to solve the primitive equations. We use a spectral resolution of T42 for most experiments, providing a grid of 64 latitude and 128 longitude cells, each of approximately \ang{2.8}$\times$\ang{2.8} in size. For higher rotation rate simulations, the decreasing Rossby deformation scale that results from a higher value of $\Omega$ demands a higher spectral resolution of T85 \citep{kaspi2015-ATMOSPHERIC, vallis2017-Atmospheric}. Each model has a vertical grid defined by 50 unevenly spaced levels up to a pressure height of \SI{0.02}{\hecto\Pa}. 

We specify a global mixed-layer slab ocean of \SI{2.5}{\m} depth without prescribed surface Q-fluxes \citep{jucker2017-Untangling}. Isca does not include a dynamical ocean model, so this allows for a closed surface energy budget with the mixed layer depth setting the thermal inertia of the slab ocean; the value is low to allow a strong response to seasonal forcing \citep{donohoe2014-Effect, jucker2019-SURFACE, liu2021-SimCloud}. An albedo of $0.2$ is used to account for the additional albedo that clouds provide \citep{liu2021-SimCloud}; the value is used in all experiments for consistency across the parameter sweep.

To represent moist convection, we use the Simple Betts-Miller (SBM) scheme described by \citet{frierson2007-Dynamics}. SBM is a quasi-equilibrium scheme \citep{betts1986-New, betts1986-Newb} that relaxes to a relative humidity profile of \qty{70}{\percent} and temperatures to a moist adiabat over a \SI{7200}{\s} timescale when convective instability occurs. This is coupled with a large-scale condensation scheme that removes excess vapour beyond saturation in each grid cell. We use \textsc{SOCRATES} \citep{edwards1996-Studies, manners2016-SOCRATES} for the radiation scheme, which provides higher complexity relative to simpler grey-radiation schemes: \textsc{SOCRATES} is a comprehensive, multi-band scheme developed by the UK Met Office which has already been used in applications beyond present-day Earth \citep{amundsen2016-UK, way2017-Resolving}. We run \textsc{SOCRATES} with the included \texttt{ga7} spectral files, providing 9 long-wave and 6 shortwave bands (standard configuration for an Earth-like model), and making use of stratospheric ozone absorption to provide a closer comparison with Earth's atmospheric temperature structure.

\subsection{Cloud Scheme and Rotation Rates}
To implement clouds in the simulations, we make use of the SimCloud scheme developed by \citet{liu2021-SimCloud}. In general, cloud schemes add significant computational overhead to models since the physics of clouds are complex, necessitating interface with multiple parts of a GCM. SimCloud aims to provide a simple diagnostic cloud scheme for examining the impacts of clouds on climate in idealised models. SimCloud diagnoses clouds by local environmental variables and only interacts with the radiation scheme. Simplifications include treating water and ice clouds both as liquid but with different effective cloud particle radii as derived from observations \citep{stubenrauch2013-Assessment}. Large-scale clouds are diagnosed using the local grid-mean relative humidity, with a freeze-dry adjustment applied to prevent the overestimation of polar clouds. Marine low-level clouds such as stratocumulus are diagnosed using the local vertical temperature profile. We employ the following options within the scheme for cloud diagnostics \citep[identical to runs from][]{liu2021-SimCloud}: a linear large-scale cloud diagnostic formula; maximum-random assumption for cloud overlap; and the Park-ELF method for low-level marine clouds. Finally, precipitation in Isca is idealised and decoupled from the cloud scheme, being driven by either large-scale condensation \citep{frierson2006-GrayRadiation} or humidity relaxation from the SBM convection scheme \citep{frierson2007-Dynamics}. This allows precipitation formation in models with or without a cloud scheme present.

We present simulations with the model using 22 different rotation rates. With Earth's rotation rate labelled $\Omega_E$, we define $\Omgstar = \Omega/\Omega_E$, which is the ratio of the planetary rotation rate compared to Earth's. $\Omgstar$ takes values between $1/128$ and $4$, extending from very slow to fast rotators, and encompassing the rotation rates of Earth, Mars and Titan. At each rotation rate, we use configurations of the model with the cloud scheme enabled and disabled to provide a comparison with a clear-sky simulation, allowing the impacts from cloud forcing to be more clearly identified. Each simulation lasts for 15 Earth years, sufficient for the large-scale dynamics of the troposphere to reach a steady-state; we discard the initial 10 as a spin-up period, and use the final five years to provide climatological averages for each experiment.

%% file: 3-results.tex
\section{Results}\label{sec:results}
\subsection{Impact of Rotation Rates on Clouds}

\begin{figure*}[t]
    \centering
    \includegraphics[width=1.0\textwidth]{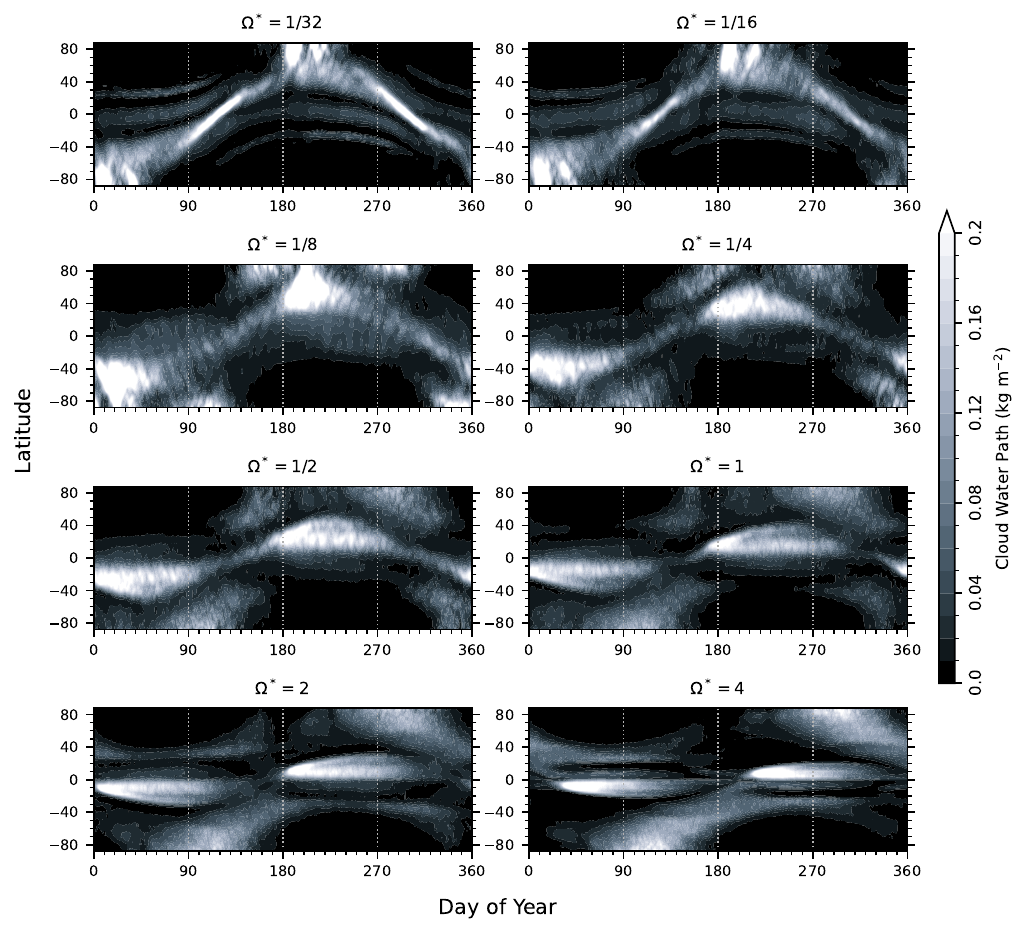}
    \caption{Global seasonal variation of zonal-mean total cloud water path (CWP) for a subset of the simulations. Lighter regions represent regions with higher CWP values.}
    \label{fig:global_cloud_cover}
\end{figure*}

A number of variables may be used to diagnose the presence of clouds in the atmosphere. One such quantity is the total cloud water path (CWP), a vertically integrated quantity defined by 
\begin{equation}\label{eq:cwp}
    \textrm{CWP} = \frac{1}{g}\int_{0}^{p_s} C w_l \;\dff p\,,
\end{equation}
where $g$ is the surface gravity, $C$ the cloud fraction and $w_l$ the liquid mixing ratio at pressure levels $p$. The CWP provides a useful proxy measure for where the most substantial clouds may be found, particularly those associated with moist, convective regions. If the value of CWP is high then we may safely conclude clouds are present, though clouds may also be present in low-valued regions, as discussed later.

Figure~\ref{fig:global_cloud_cover} shows the global variation of CWP throughout the seasonal cycle across a range of rotation rates. It is immediately clear that the average spatial distribution of CWP varies significantly with both rotation rate and time of year, with some key identifiable trends. Starting with $\Omgstar=1$, the most Earth-like of the experiments, there is a very clear band of high CWP that straddles the equatorial region during near-solstice periods (approximately day 0 and 180) and undergoes a weak latitudinal migration during the seasonal cycle. This is the ITCZ, where the convergence and ascent of moist, warm air is highly conducive to the formation of clouds \citep{pierrehumbert2010-Principles}.

Considering the more slowly rotating systems, the ITCZ-like cloud band broadens in latitudinal extent but also develops a strongly seasonally migratory nature, extending well into what are considered extratropical latitudes on Earth. At $1/16<\Omgstar<1/4$ the ITCZ cloud band transitions from a near-stationary regime to one wherein its seasonal migration reaches the polar regions near solstice, consistent with idealised simulations of Titan \citep{mitchell2006-Dynamics}. The equinoxes (approximately days 90 and 270) represent a transitional state where the major cloud band migrates rapidly across the equatorial region. High values of CWP are predominantly found in the summer hemisphere, with the winter hemisphere being mostly cloud-free. This behaviour in the solsticial period becomes ever clearer in the slowest rotating models, hinting at the development of asymptotic behaviour.

At rotation rates faster than Earth's, a different regime occurs. The tropical cloud band appears to become weaker and much more restricted in latitudinal extent, agreeing with previous results \citep{yang2013-STABILIZING}. At the same time, new bands of high/low CWP occur at higher latitudes. Relating the position of these bands to Earth's climate suggests they are associated with extratropical storm tracks that develop in an increasingly baroclinic atmosphere \citep{kaspi2013-Role, shaw2016-Storm}; in addition to cloud-free subtropical regions. These additional bands are most clearly developed near the equinoxes, which contrasts with the slowly rotating systems where the greatest distinction is apparent near the solstices.

\begin{figure*}[tb]
    \centering
    \includegraphics[width=1.0\textwidth]{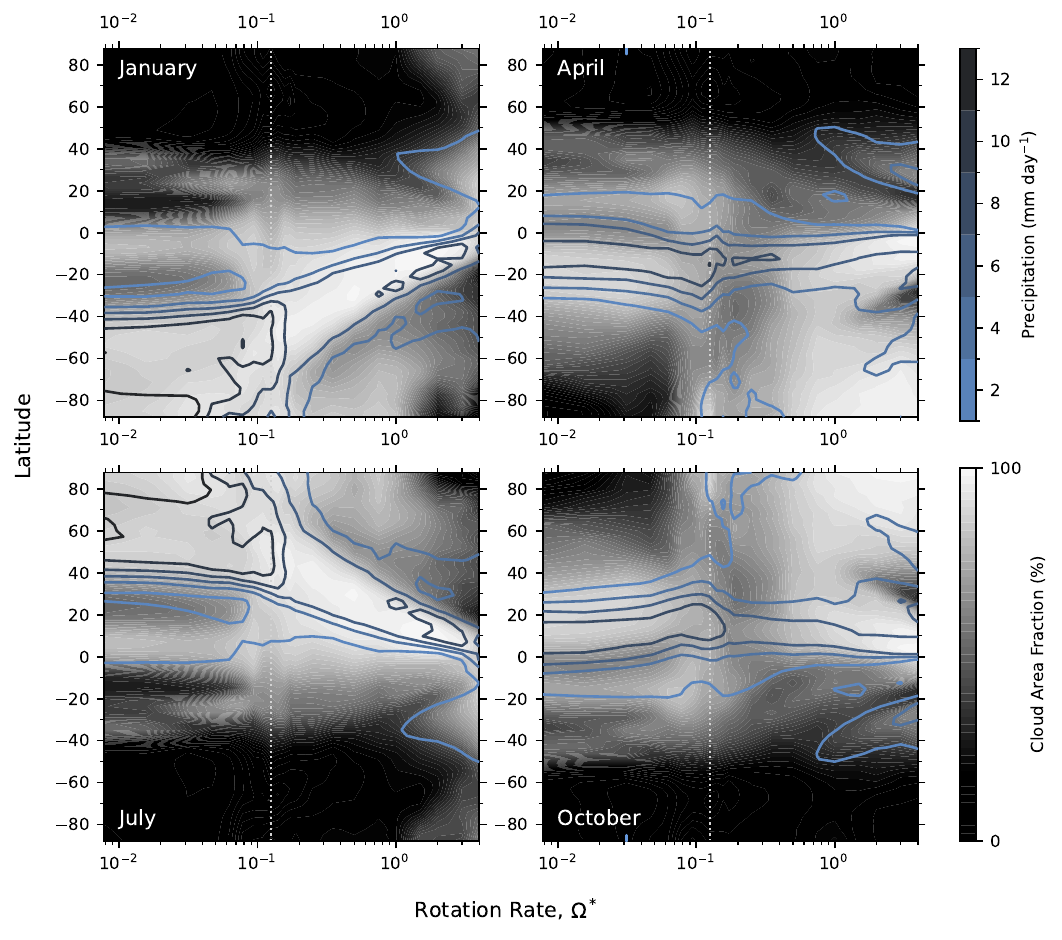}
    \caption{Variation of cloud area fraction and precipitation rate with rotation rate. Filled contours show percentage coverage of cloud viewed from space, with lighter regions being more cloudy; blue lines represent contours of increasing precipitation. The vertical dotted line represents a rotation rate of $\Omgstar=1/8$.}
    \label{fig:cloud_rotation_seasonal}
\end{figure*}

Whilst Figure~\ref{fig:global_cloud_cover} gives us a snapshot of a select subset of experiments, it does not show the transition that occurs over the full range of rotation rates. Figure~\ref{fig:cloud_rotation_seasonal} shows the zonally averaged behaviour for all rotation rates. In addition to precipitation, we also show the overall cloud area fraction as viewed from space; clouds at any height in the atmosphere contribute, with the maximum-random overlap assumption in the cloud scheme applied.

Figure~\ref{fig:cloud_rotation_seasonal} shows the behaviour of the main ITCZ-like cloud band described above, traced out by high contours of precipitation; the poleward extent in the solsticial period at low rotation rates is very clear. Overall, the total cloud area fraction suggests a much greater abundance of clouds than previously suggested by the CWP alone. This is because CWP is a vertical integral of cloud water so suppresses the influence of certain clouds like shallow marine stratocumulus. This suggests that there are regions aside from those identified qualitatively in Figure~\ref{fig:global_cloud_cover} that are cloudy and consequently contribute to the radiation balance of the climate system. One clear behaviour nonetheless is the absence of any clouds in the polar regions of the winter hemisphere at all but the fastest rotation rates. This implies the existence of a consistent mechanism that suppresses cloud formation at a wide range of rotation rates. At the highest rotation rates, there is the development of a cloudy region in the high latitudes around the autumnal equinox. The high cloud area fraction is co-located with high precipitation rates---it is possible this is a manifestation of the storm tracks, but it may also be an over-estimation of polar clouds despite the freeze-dry adjustment in the SimCloud scheme, as identified in \citet{liu2021-SimCloud}.

\begin{figure*}[tb]
    \centering
    \includegraphics[width=1.0\textwidth]{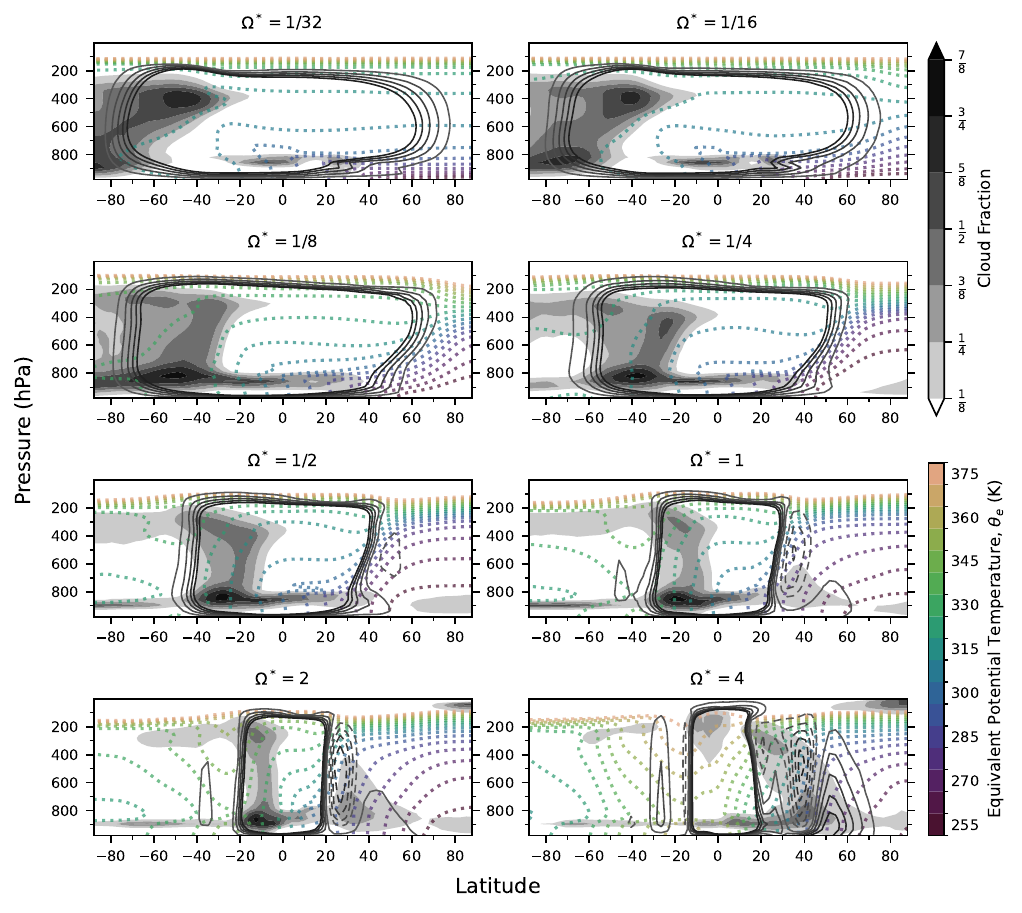}
    \caption{Zonally-averaged relationship between cloud location, atmospheric circulation and convective stability at a range of rotation rates, averaged across January. Filled grey contours show grid cell cloud fraction. Mass streamfunction $\psi$ is denoted by solid contours at \SI{2}{\percent} intervals up to \SI{10}{\percent} of the maximum value for each subplot; \SI{10}{\percent} is found at inner edge of main circulation cells. Solid lines denote clockwise circulation; dashed lines denote anti-clockwise. The maximum values are, respectively, 18.1, 16.9, 10.9, 10.3, 6.5, 5.1, 3.1, \SI{1.0e11}{\kg\per\s}. Coloured dotted contours show the equivalent potential temperature surfaces in the lower atmosphere.}
    \label{fig:cf_psi_ept_DJF}
\end{figure*}

Having shown that the spatial distribution of seasonal clouds on a terrestrial aquaplanet is strongly influenced by the rotation rate, we turn to understanding some of the mechanisms driving these changes. Figure~\ref{fig:cf_psi_ept_DJF} links the zonally-averaged distribution in cloud fraction to diagnostic variables related to the broader atmospheric state. Firstly, the meridional mass streamfunction, $\psi$, is calculated as 
\begin{equation}\label{eq:massstreamfunction}
    \psi\left(\vartheta,p\right) = \frac{2\pi a}{g} \cos\vartheta \int v\left(\vartheta,p\right) \;\dff p \,,
\end{equation}
where $\vartheta$ is the latitude, $a$ the planetary radius and $v$ the meridional wind. This quantity gives us an understanding of the meridional atmospheric circulation by highlighting where regions of ascending and descending air occur. On Earth, analysis of $\psi$ indicates the presence of the Hadley, Ferrel and polar cells, which delineate the principal circulation regimes of the terrestrial atmosphere. The peak streamfunction magnitude will vary between models at different rotation rates, as shown by \citet{kaspi2015-ATMOSPHERIC}. 

The second diagnostic variable is the equivalent potential temperature $\theta_e$. For any given air parcel, the equivalent potential temperature is conserved with vertical motion and phase changes of the condensing species. The precise formula for $\theta_e$ is complicated \citep{paluch1979-Entrainment, emanuel1994-Atmospheric} so we use an approximated form \citep{stull1988-Introduction} wherein
\begin{equation}\label{eq:theta_e}
    \theta_e \approx \left(T+r\frac{L_v}{c_p}\right)\left(\frac{p_s}{p}\right)^\kappa \,,
\end{equation}
where $T$ and $r$ are temperature and vapour mixing ratio respectively. Constants include enthalpy of vapourisation $L_v$, gas heat capacity (at constant pressure) $c_p$, surface pressure $p_s$, and $\kappa$, defined by the ratio of moist to dry gas constants ($R_v/R_d$). The term $r$ is substituted for specific humidity $q$ (a model output) since in an Earth-like climate $q\approx r$ \citep{vallis2017-Atmospheric}. Equivalent potential temperature is a useful quantity since it allows us to diagnose the stability of the atmosphere with respect to convection---clouds are strongly linked to vertical mixing in moist regions. If $\theta_e$ increases with height ($\dfp\theta_e/\dfp z > 0$) then the atmosphere is stably stratified and vertical motions are suppressed. Conversely, if $\dfp\theta_e/\dfp z < 0$, the air is unstable to vertical perturbations and convection occurs. Moist parcels of air lifted by convection and thus cooled contribute to cloud formation.

Figure~\ref{fig:cf_psi_ept_DJF} shows the January average for eight different rotation rates, reflecting the climate in northern hemisphere (NH) winter and southern hemisphere (SH) summer. The variation of the main large-scale meridional overturning circulation pattern diagnosed by the mass streamfunction, the Hadley cell, is clear through all the rotation rates: in the slowly rotating cases the Hadley cell extends into the high mid-latitudes with ascending/descending branches in the summer/winter hemispheres, a type of circulation observed on Titan \citep[e.g.,][]{mitchell2016-Climate}. In these cases, the bulk of the clouds occur in the mid-latitude and polar regions throughout the depth of the summer troposphere: at higher levels, extending up to \SI{200}{\hecto\Pa}, the cloud fraction is highest on the equatorward flank of the rising branch of the Hadley cell, whereas at lower levels the cloudiest regions occur at higher latitudes, producing a somewhat tilted structure. The high cloud fraction found near the ascending branch of the Hadley cell can be understood through the gradient in $\theta_e$, which does not increase strongly with height and is close to zero in localised regions. This indicates convection, which saturates and condenses out excess moisture, thus contributing to cloud formation. Low-level stratocumulus clouds are additionally under the influence of boundary layer dynamics and instabilities, which may explain their more cosmopolitan distribution compared to the large-scale clouds.

In contrast to the summer cloudiness, the winter hemisphere is comparatively devoid of any large-scale cloud formation in the slowly rotating cases and there is a clear positive gradient of $\theta_e$ with height. The only exception to an otherwise cloud-free hemisphere is the limited presence of low-level clouds in the low latitudes. These large-scale distributions can be explained by considering the large-scale circulation together with the atmospheric stability. In the winter hemisphere air parcels along the descending branch of the Hadley cell warm adiabatically. Cloud formation is effectively inhibited by the downwelling air, which manifests as a consistent gradient in $\theta_e$, which increases with height.

As rotation rate increases, the latitudinal extent of the Hadley cell decreases, and with it moves the location of peak cloud fraction. At Earth's rotation rate $\Omgstar=1$, the Hadley cell is much reduced and there is development of a significant second, thermally indirect cell in the mass streamfunction, corresponding to the Ferrel cell. Similar trends in the development of large-scale circulation at near-solstice periods can be seen in the study by \citet{faulk2017-Effects}, supporting the link between large-scale circulation, convective stability and cloud formation. Whilst low-level clouds are more widespread at Earth's rotation rate, the large-scale clouds continue to align with regions of vertical motion where there is not a clear positive gradient of $\theta_e$ with height. In addition to the deep cloud structure in the tropics, a secondary region of deep cloud in the winter mid-latitudes also begins to develop, extending up to \SI{500}{\hecto\Pa} in height. This becomes the dominant cloud structure as the rotation rate increases to higher values and is an indicator of extratropical storm track activity.

At the highest rotation rates the dynamical situation becomes more complex. In slowly rotating systems energy transport is dominated by the mean-flow, reflected by the global Hadley circulation. At higher rotation rates energy transport becomes eddy-dominated \citep{kaspi2015-ATMOSPHERIC} and secondary circulation cells form. As $\Omega$ increases the eddy length scale decreases \citep{kaspi2015-ATMOSPHERIC, showman2015-THREEDIMENSIONAL, rhines1979-Geostrophic}, necessitating the formation of multiple smaller circulation cells and eddy-driven jets \citep{chemke2015-Poleward, wang2018-Comparative}, explaining the structure in $\psi$ at the highest rotation rates. This reflects the highly baroclinic nature of the winter troposphere in the mid-latitude regions, wherein extratropical cyclones are prominent. These regions of low pressure promote cloud formation as warm, moist, poleward-flowing air is lifted along fronts, leading to substantial large-scale cloud formation. The preferred locations of these systems trace out the storm tracks, seen as the increased cloud fraction in the mid-latitudes. As before, regions of high cloud fraction often align with regions where $\dfp\theta_e/\dfp z$ is not strongly positive. Whilst clouds become more prevalent in the extratropics, the deep convective cloud band formerly present in the slower rotation rates dissipates, likely as a result of a weakening Hadley cell where convective updraughts are less significant. In contrast to the slowly rotating cases, it is now the summer hemisphere that is almost devoid of clouds, again with the exception of the low-level clouds which likely result from boundary layer turbulence.

\begin{figure}[tb]
    \centering
    \includegraphics[width=1.0\columnwidth]{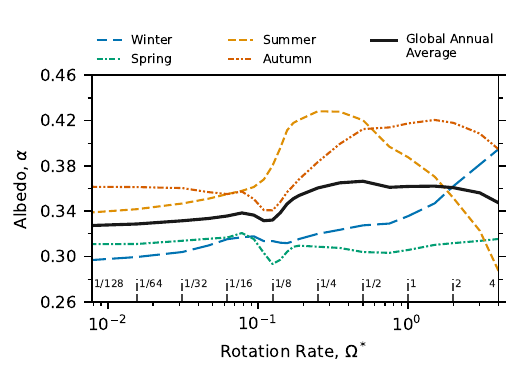}
    \caption{Planetary albedo at varying rotation rates, split by hemisphere. Coloured lines are formed by the average of the corresponding season in each hemisphere: Winter (Northern hemisphere DJF, Southern hemisphere JJA); Spring (NH MAM, SH SON) and so on. The solid black line represents the resulting global average over both hemispheres.}
    \label{fig:albedo_seasonal_variation}
\end{figure}

Figure~\ref{fig:albedo_seasonal_variation} shows the overall impact of the presence of clouds, accounting for changes in spatial distribution with rotation rate, on the planetary albedo. Albedo is a measure of the incoming versus outgoing shortwave radiation and is an important quantity for determining the overall radiation budget of a planet. Both seasonal hemispheric and annual global averages are calculated for planetary albedo, produced by considering the top-of-atmosphere (TOA) shortwave radiation
\begin{equation}
    \alpha = \frac{S_{\uparrow,\textrm{TOA}}}{S_{\downarrow,\textrm{TOA}}}\;,
\end{equation}
where $S_\downarrow$ and $S_\uparrow$ are the incoming and outgoing shortwave radiative fluxes. Following calculation of the TOA albedo $\alpha$ on a grid-cell basis, an area weighting is applied before  averaging spatially and temporally. 

Given each simulation uses a surface albedo $\alpha=0.2$, it is immediately clear that clouds act to increase the global TOA albedo in every season, with values varying around $\alpha \approx 0.35$. This behaviour is expected since clouds reflect the incoming shortwave radiation back to space, increasing $S_\uparrow$. At the slowest rotation rates the global albedo appears to tend asymptotically towards a value of $\alpha \approx 0.33$, whereas at high rotation rates it decreases again from the maximum value. This perhaps unexpectedly non-monotonic behaviour can be understood by linking to the analysis of large-scale cloud formation and atmospheric circulation.

Considering seasonal averages, the variation with rotation becomes more pronounced. The summer season displays the highest albedo peak of any season throughout the parameter space, which may be explained by the presence of the deep cloud band along the ascending Hadley cell branch. Near the peak value, found near $\Omgstar\approx 1/4$, this dominant cloud band is shown by Figure~\ref{fig:cf_psi_ept_DJF} to be centred around \ang{30}S; during the DJF period the solar zenith angle will be very high and as such clouds have ample ability to impact the albedo. Simulations at these rotation rates also display strong low-level stratocumulus decks, which are known for having an important impact on energy balance and albedo \citep{slingo1990-Sensitivity, hartmann1992-Effect}. At slower rotation rates the Hadley cell develops a fully pole-to-pole circulation, with peak cloud formation occurring closer to the summer pole.  Conversely, at higher rotation rates, changes to the Hadley cell circulation result in less significant cloud formation in the summer hemisphere. 

The broad peak in the autumnal albedo can plausibly be explained by considering the importance of eddies, which is shown by \citet{kaspi2015-ATMOSPHERIC}, in terms of energy flux, to peak around Earth's rotation rate. Eddies imply the existence of extratropical cyclones and their associated clouds. At slower rotation rates the extratropical regions shrink, whilst at higher rotation rates the eddy length scale decreases, which also acts to shrink the cloud formation regions. The cloud formation region is further restricted by the development of convergent subtropical areas that supress cloud formation (analogous to Hadley/Ferrel cell boundary on Earth).

Overall the winter hemisphere displays a relatively low albedo, which supports the notion that a high solar zenith angle results in clouds having a smaller impact on the shortwave radiation budget. Nonetheless, there is a marked increase at the fastest rotation rates, which may be attributed to cloud formation associated with storm track activity in the extratropics, as with the autumnal case.

Finally, spring appears to consistently have low albedo values, which reflects a relatively cloud-free season during which the solar zenith angle is still relatively high. Comparing Figures~\ref{fig:global_cloud_cover} and \ref{fig:cloud_rotation_seasonal}, one may deduce that the limited cloud present does not have significant CWP, indicating that it is most likely low-level stratocumulus driven more by the boundary layer, rather than by the large-scale circulation, explaining the lack of variation with rotation rate.

The cumulative effect of the seasonal albedos produces an unexpected variation in the annually averaged global albedo, which shows a clear non-monotonic behaviour. Aside from the key trends at the extrema and the peak value, also present is a distinctive kink that occurs around $\Omgstar=1/8$ where the albedo has a local minimum---a feature clear in all seasons except summer. This rotation rate appears to have some significance with respect to atmospheric dynamics: it represents the transition between a single global Hadley cell that extends fully from pole to pole to a more restricted form where additional meridional circulation cells (i.e., Ferrel cells) are present. This transition at $\Omgstar \sim 1/8$ also appears in the results of \citet{kaspi2015-ATMOSPHERIC} as the rotation rate at which mean meridional heat transport via the large-scale circulation is overtaken by poleward eddy heat transport. The transitional range of $\Omega$ between these two dynamical regimes appears to briefly suppress cloud formation, perhaps as neither mechanism that normally leads to cloud formation---overturning circulation or baroclinic instabilty---is dominant.

\subsection{Impact of Clouds on Climate and Seasonality}\label{sec:results-seaonality}

Previous studies, ignoring clouds, have shown the impact of rotation rates on climate and seasonality, including the ITCZ extent, seasonal amplitude, and latitudinal extent of the maximum temperature \citep[e.g.,][]{faulk2017-Effects,geen2019-Processes,guendelman2022-Key}. Building on these insights, we investigate how clouds impact climate and seasonality under varying rotation rates.

\begin{figure*}[htb]
   \centering
   \includegraphics[width=1\textwidth]{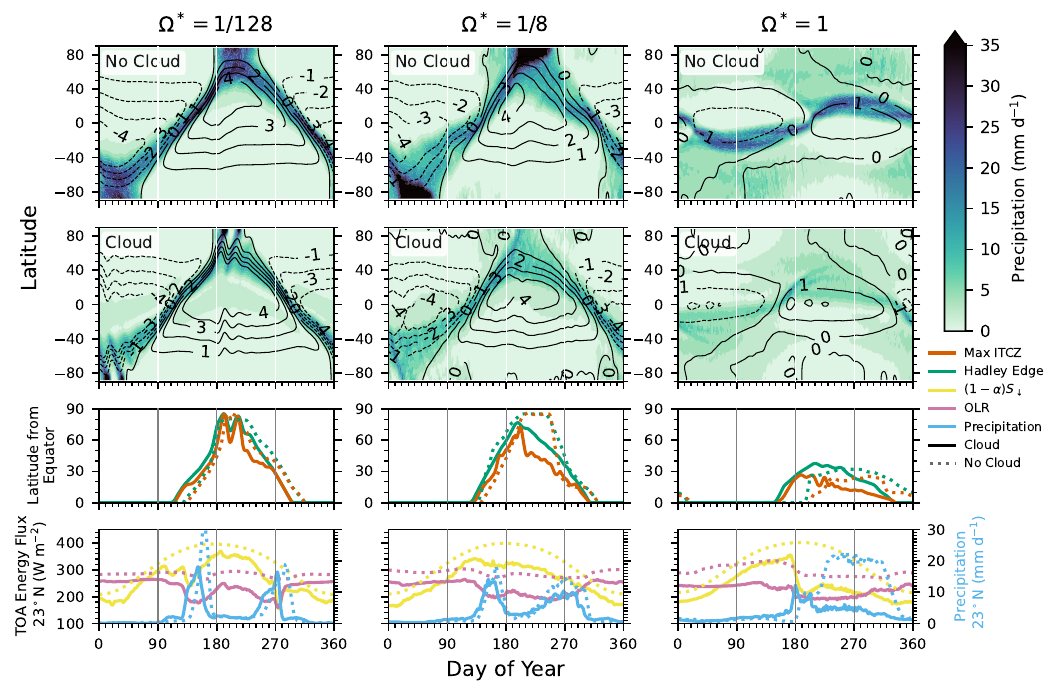}
   \caption{Variation of precipitation and mass streamfunction for different relative rotation rates: $\Omgstar=$ 1/128 (left), 1/8 (middle), and 1 (right) for simulations without clouds (upper row) and simulations with clouds (middle row). The filled contours depict precipitation (mm d$^{-1}$). The black contours show zonal-mean mass streamfunction at the pressure level of its maximum value with inline labels of values in units of \SI{e11}{\kg\per\s}, and all panels share the common contour interval of \SI{1}{\kg\per\s}. The third row displays the inner edge of the winter Hadley cell in the summer hemisphere, and the maximum zonal mean precipitation (Max ITCZ) in the Northern Hemisphere. The solid curves show results from the simulations with clouds, while the dotted curves correspond to the simulations without clouds. The ITCZ is defined as the latitude of maximum zonal-mean precipitation and it must be associated with the ascending branch of the Hadley circulation; the edge of the Hadley cell is defined as the latitude where the streamfunction, taken at the pressure level of its maximum value, reaches $5\%$ of that maximum value in the summer hemisphere. The bottom row illustrates the changes in the zonal-mean TOA energy fluxes (left y-axis) and precipitation (right y-axis) over time at a latitude of 23$^\circ$N. The yellow lines represent the net shortwave radiation at TOA ($1-\alpha)S\downarrow$, while the pink lines represent outgoing longwave radiation (OLR). The blue lines depict the zonal-mean precipitation. The vertical lines represent day of vernal equinox (day 90), northern summer solstice (day 180), and autumnal equinox (day 270), and day 0 is northern winter solstice. The presented data are daily averages. The contours of streamfunction in the first two rows and the curves in the third row are smoothed by a Gaussian filter with a standard deviation of $\sigma = 3$ days.}
   \label{fig:ITCZ}
\end{figure*}

Figure~\ref{fig:ITCZ} displays the zonal mean precipitation and mass streamfunction for both simulations without clouds and simulations with clouds in the top two rows. Our simulations show that, regardless of the presence of clouds, the extent of the Hadley cell and ITCZ decreases as rotation rates increase. However, simulations with clouds exhibit lower maximum precipitation over the year, indicating that clouds may suppress ITCZ precipitation, while the streamfunctions are stronger than in the no-clouds case. In the third row, we present how the inner edge of the winter Hadley cell in the summer hemisphere, and the ITCZ change seasonally in the NH. In no-cloud simulations, both the Hadley cell and ITCZ tend to remain at the maximum latitude for longer after reaching the summer maximum latitude, compared to the simulations with clouds. In the bottom row, the zonal mean TOA net shortwave radiation ($(1-\alpha)S_{\downarrow}$) and outgoing longwave radiation (OLR) at \ang{23} latitude are shown separately, along with the change of zonal mean precipitation.

Our no-cloud simulations exhibit a pattern qualitatively consistent with the findings of \citet{faulk2017-Effects} regarding slow-rotating planets ($\Omgstar \sim 1/8$), with some differences in the latitudes of ITCZ and Hadley Cell extent. According to \citet{faulk2017-Effects}, such planets have global Hadley cells and are essentially `all tropics' planets, with the ITCZ remaining at around $\sim\,$\ang{70}. Our no-cloud simulations have the ITCZ extending closer to the summer pole for slow rotation rates ($\Omgstar<1/8$). For the no-clouds case with $\Omgstar=1/8$, the maximum precipitation remains near the summer pole for about 50 days, along with the ascending branch of the Hadley cell, and the highest precipitation values exceed \SI{35}{\mm\per\day}.

We locate the ITCZ at the latitude of the maximum zonal-mean precipitation. However, in the slowest rotating case ($\Omgstar=1/128$), a transient maximum in zonal-mean precipitation is observed at the summer pole (days 180--230). Therefore, we choose the other precipitation maximum at a lower latitude ($<\,$\ang{80}), which is associated with the ascending branch of the Hadley cell, as the ITCZ. The transient polar precipitation maximum is also found by \citet{faulk2017-Effects} in the simulations with $\Omgstar<1/6$, while in our no-clouds simulations, it only appears in the $\Omgstar=1/128$ case. In this simulation, the extent of the ITCZ and Hadley cell is lower than the case with $\Omgstar=1/8$ and the ITCZ is located southward of the ascending branch of the Hadley cell.

For our simulations with clouds, the highest NH mean precipitation over the year occurs in early summer for all cases, along with the maximum latitude of the ITCZ and Hadley cell. In simulations with $\Omgstar=1/8$, the precipitation rapidly decreases after reaching its peak. Interestingly, as the rotation rate decreases from $\Omgstar=1/8$, a second peak of zonal-mean precipitation occurring in the late summer begins to develop, along with increases in the latitudes of the ITCZ and Hadley cell edge. At the same time, the absolute peak value of precipitation increases monotonically as the rotation rate decreases from $1/8$ to $1/128$. The double-peak precipitation pattern might relate to the findings on polar precipitation for high-obliquity cases reported by \citet{lobo2020-Atmospheric}, where the peaks were attributed to the moisture storage term, which is usually negligible in Earth’s climate.

The relationship between TOA radiative energy imbalance and precipitation at a latitude of \ang{23} is illustrative. In both cloudy and cloud-free scenarios, precipitation initiation coincides with the moment when the net energy becomes positive. At slower rotation rates ($\Omega^* = 1/128, 1/8$), the most intense precipitation occurs when the ITCZ passes, twice a year. For Earth's rotation rate, the ITCZ remains near the \ang{23} latitude for several months, resulting in prolonged and intense precipitation. In the no-cloud simulations, the net shortwave radiation is solely determined by the solar zenith angle, and OLR remains relatively constant over the course of a year, being minimally influenced by rotation rate. However, the intensity of precipitation varies with rotation rate, indicating that strong precipitation is primarily governed by atmospheric circulation rather than local TOA radiative energy. On the other hand, in simulations with clouds, precipitation strongly impacts the local TOA radiation: both shortwave and longwave radiation decrease with increasing precipitation, reflecting shortwave and longwave cloud radiative forcing. Overall, the presence of clouds suppresses precipitation, and at all three rotation rates shown shifts the peak precipitation to earlier in the season.

\begin{figure}[htb]
    \centering
    \includegraphics[width=\columnwidth]{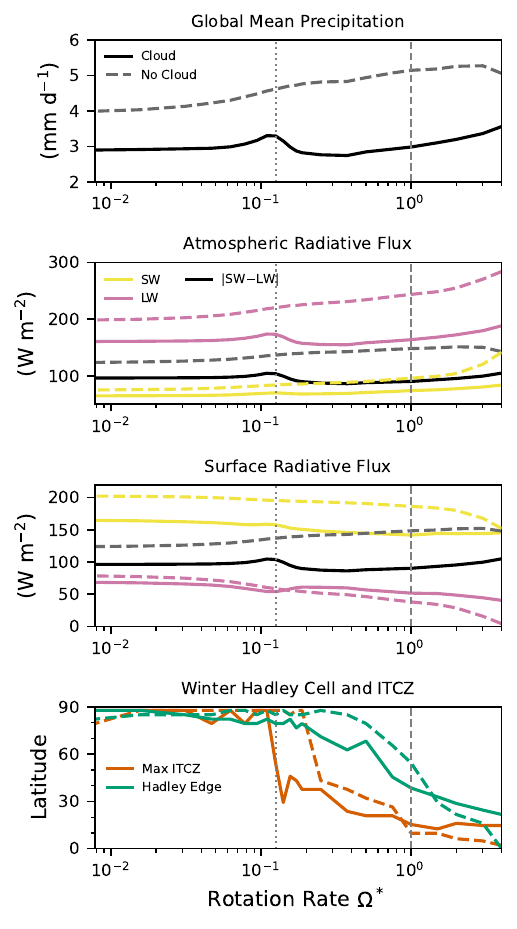}
    \caption{Top: variation of global and annual mean precipitation with rotation rate. Second panel: the shortwave radiative flux absorbed by the atmosphere (yellow), the net longwave radiative loss from the atmosphere (pink), and the net atmospheric radiative energy input (black). Third panel: the shortwave radiative flux absorbed by the surface (yellow), the net longwave radiative loss of the surface (pink), and the net surface radiative flux. Bottom: the highest latitude during the year of the maximum zonal mean precipitation (Max ITCZ) and the inner edge of the winter Hadley cell in the summer hemisphere. All panels show simulations with clouds (solid lines) and without clouds (dashed lines). The vertical lines represent rotation rates of $\Omgstar$ = 1/8 and 1.}
    \label{fig:lat-rotation}
\end{figure}

The global average precipitation in both cases, with and without clouds, changes with the rotation rate. The upper panel of Figure~\ref{fig:lat-rotation} shows the global annual mean precipitation. The atmospheric net radiative energy is presented in the middle panel, which aligns with the precipitation variations \citep{ogorman2008-Hydrological}. The simulations without clouds have overall higher global annual mean precipitation at any rotation rate, and the precipitation increases with rotation rate monotonically. However, when clouds are present, a distinctive kink appears in the precipitation curve around $\Omgstar = 1/8$, corresponding to the kink in albedo presented in Figure~\ref{fig:albedo_seasonal_variation}. Regarding the atmospheric energy budget, clouds tend to decrease the absorbed shortwave radiation flux in the atmosphere, but they also decrease longwave radiative loss (second panel of Figure~\ref{fig:lat-rotation}). Our results suggest that the longwave radiative forcing from clouds is the dominant effect over the shortwave radiative forcing at all rotation rates. Examining the surface energy budget further (the third panel of Figure~\ref{fig:lat-rotation}), we observe that clouds decrease the shortwave radiation flux; they also decrease longwave radiative cooling when $\Omgstar<1/8$ but increase it when $\Omgstar>1/8$. In contrast to the atmospheric radiative budget, the shortwave surface forcing from clouds dominates over the longwave forcing at all but the highest rotation rates.

The dependence of the winter Hadley cell extent and ITCZ on the rotation rate is shown in the bottom panel of Figure~\ref{fig:lat-rotation}. When $\Omgstar<1/8$, there is no significant difference between non-cloud and cloudy cases, while at around $\Omgstar=1/6$, simulations without clouds exhibit more extended Hadley cells and therefore ITCZ latitude than those with clouds. As the rotation rate increases, the discrepancy between the two cases diminishes and eventually reverses sign around the Earth's rotation rate. Note that the influence of clouds on ITCZ latitude is metric-dependent. Here, the extent of ITCZ is defined by the latitude of the maximum zonal-mean precipitation in the NH, and the decreases of ITCZ extent caused by clouds appear within $1/6<\Omgstar<1$. If we define it instead by the maximum latitude of the ITCZ over the year (the peak of the green curve shown in the bottom panels of Figure~\ref{fig:ITCZ}), the decrease appears within $3/16<\Omgstar<3/8$. Regardless of the metric used, the decrease in ITCZ extent due to clouds only occurs at intermediate rotation rates.

\begin{figure}[tb]
   \centering
   \includegraphics[width=\columnwidth]{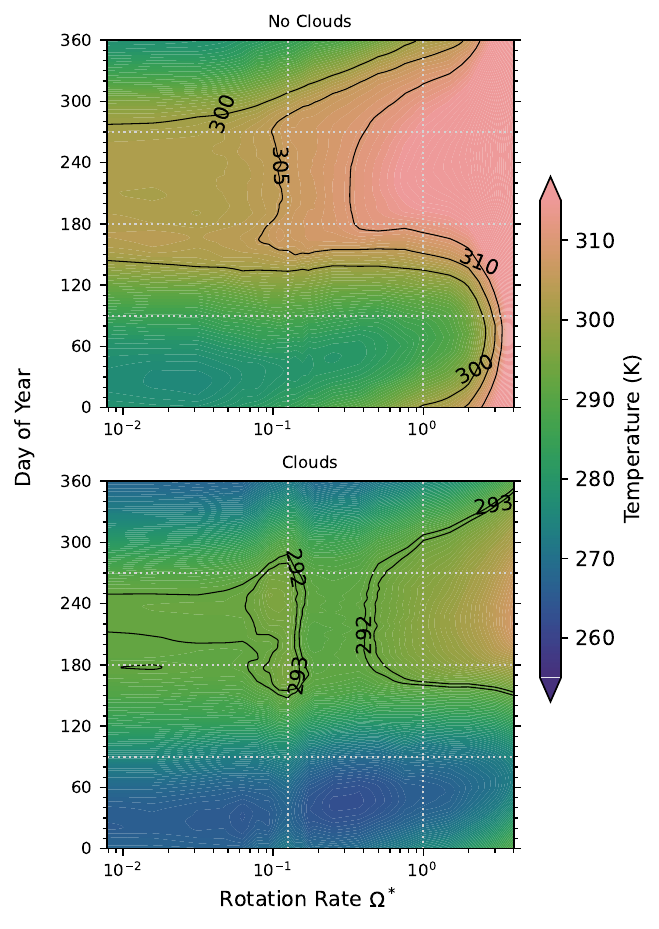}
   \caption{Daily variation of the Northern Hemisphere (NH) mean surface temperature with different rotation rates for simulations without (upper) and with clouds (lower). The vertical lines represent rotation rates of $\Omgstar$ = 1/8 and 1. The horizontal lines represent day of vernal equinox (day 90), northern summer solstice (day 180), and autumnal equinox (day 270), and day 0 is northern winter solstice. To highlight the continuous structure in the upper panel and the distinctive maxima in the lower panel, two open contour lines are also shown.}
   \label{fig:T_diff1}
\end{figure}

Climate seasonality can be considered as the deviation from the annual mean climate or deviations from hemispheric symmetry for an aquaplanet \citep{guendelman2022-Key}. There are various methods for quantifying these deviations, and we will focus on temperature first. The daily variation of the NH mean surface temperature as a function of the rotation rate is shown in Figure~\ref{fig:T_diff1}. The temperatures in the no-clouds cases are generally higher than those in the cases with clouds. This is because the presence of clouds raises the overall albedo. 

For simulations without clouds, the NH average temperature reaches the maximum in early summer and decreases quickly afterwards when $\Omgstar<1/8$. The local temperature peaks at the time when the ITCZ is passing. Locally, there may be a second peak temperature during the summer occurring when the ITCZ is migrating southward, but it is not as high as the first one. When $1/8<\Omgstar<1$, the temperature remains constant during the summer for approximately 100 days. When the rotation rate is that of Earth, the summer temperature starts to shift relative to the season, and it reaches the maximum in the middle of the summer, which is consistent with our experience on Earth.

For simulations with clouds, surface temperatures are generally lower, consistent with the energy balance arguments above. For fast-rotating cases ($\Omgstar>1/8$), the overall patterns are similar to those without clouds. However, a unique characteristic is observed at around $\Omgstar = 1/8$, where the temperature exhibits a two-peaked feature with maximum values occurring in both early and late summer. The maximum temperatures are also slightly higher than those at somewhat faster rotation rates, which is in contrast to the behaviour of the simulations without clouds. Finally, for rotation rates below approximately $\Omgstar = 1/8 $, there is only one peak temperature, occurring in mid-late summer.

\begin{figure}[tb]
   \centering
   \includegraphics[width=\columnwidth]{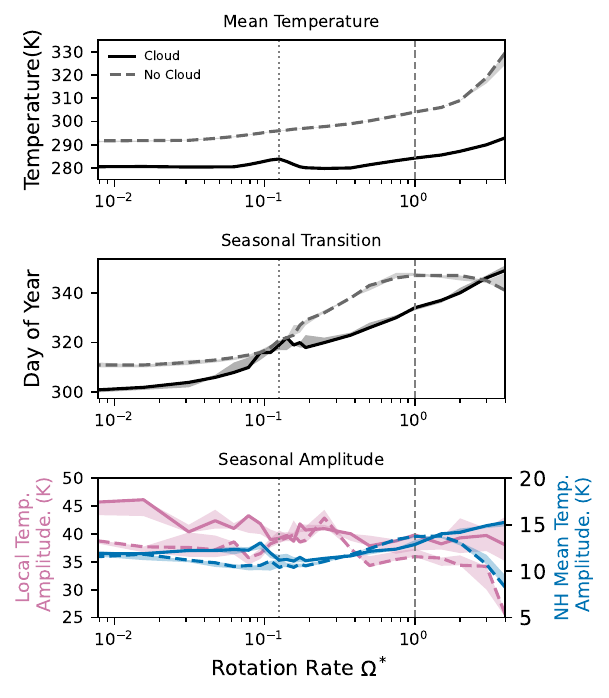}
   \caption{Top: variation of the change in the Northern Hemisphere (NH) mean surface temperature ($\overline{\langle T\rangle_\textrm{NH}}$) with rotation rate. The shaded areas represent the uncertainty range using the last five years; the uncertainty is too small to be distinguished for most of rotation rates. Middle: variation of the day when the NH mean surface temperature equals the global annual mean surface temperature after summer ($\langle T\rangle_\textrm{NH}=\overline{\langle T\rangle_\textrm{glb}}$) with rotation rate, which illustrates the lag of the seasonal transition after the autumnal equinox (day 270). Bottom: dependence of the amplitude of the seasonal temperature variation with rotation rate. The pink curve represents the maximum local temperature variation (max($T-\overline{T}$)), and values refer to the left $y$-axis. The blue curve represents the difference between the maximum NH mean temperature and the global annual mean temperature  (max$\left(\langle T\rangle_\textrm{NH}-\overline{\langle T\rangle_\textrm{glb}}\right)$ ), and corresponds to the right $y$-axis. All panels show simulations with clouds (solid lines) and without clouds (dashed lines). The vertical lines represent rotation rates of $\Omgstar$ = 1/8 and 1.}
   \label{fig:Seasonal_Transition}
\end{figure}

We further analyse three variables that represent the NH mean temperature ($\overline{\langle T\rangle_\textrm{NH}}$), the phase lag of seasonal transition from summer to winter, and seasonal amplitude for both maximum local variation ($\textrm{max}\left(T-\overline{T}\right)$) and global variation ($\textrm{max}\left(\langle T\rangle_\textrm{NH}-\overline{\langle T\rangle_\textrm{glb}}\right)$), as shown in Figure~\ref{fig:Seasonal_Transition}. The top panel of the figure shows that the NH annual mean temperature changes with rotation rate for both cases with and without clouds. Overall, the no-clouds cases have a higher mean temperature at any rotation rate, which is consistent with the global annual mean precipitation. In the absence of clouds, the temperature increases monotonically with the rotation rate, but in the presence of clouds, a kink occurs at around $\Omgstar = 1/8$, which is associated with the kink of albedo and precipitation (Figure~\ref{fig:albedo_seasonal_variation}, \ref{fig:lat-rotation} \& \ref{fig:T_diff1}).

The day of vernal equinox is day 270, but the day when the NH mean surface temperature equals the global annual mean surface temperature occurs later than equinox due to the thermal inertia, as shown in the middle panel of Figure~\ref{fig:Seasonal_Transition}. As the rotation rate increases, simulations both with and without clouds experience increasing seasonal transition lags, meaning that the phase lag with respect to the radiative forcing becomes greater. The presence of clouds overall decreases the phase lag and causes Spring temperatures to arrive earlier than in the no-clouds cases. It is worth noting that the lower albedo in the no-clouds experiments results in a higher effective emission temperature, which could, under certain approximations, decrease the radiative timescale\citep{mitchell_effects_2014,ohno2019_atmospheres,guendelman2022-Key,tan2022-Weak}. We might expect a shorter radiative timescale for the cases without clouds, corresponding to a smaller phase lag and larger variation. This expectation contradicts our simulation results, prompting a need for a more detailed calculation with our realistic radiative scheme. Below, we analyse how other factors influence the seasonal transition.

According to the bottom panel of Figure~\ref{fig:Seasonal_Transition}, the amplitude of seasonal variations of temperature does not vary substantially with rotation rate, which is consistent with the findings of \citet{guendelman2022-Key} (cf. their Fig. 3f). Despite these unclear trends, differences between the clouds and no-clouds cases becomes significant when the rotation rate is relatively high ($\Omgstar>1$) and low ($\Omgstar<1/8$).

\citet{guendelman2022-Key} pointed out a key factor controlling seasonal transitions is atmospheric heat transport efficiency. They concluded that the amplitude of temperature variation decreases as heat transport increases. Their results show that, as rotation rate increases, heat transport becomes less efficient, leading to an increase in the equator-to-pole temperature difference. This observation aligns with previous work on aseasonal planets, including simulations without clouds \citep[e.g.,][]{showman2015-THREEDIMENSIONAL,cox2021_radiative} and simulations with clouds \citep[e.g.,][]{komacek2019-Atmospheric,liu2017-effect}. In our simulations without clouds, the annual mean equator-to-pole temperature difference increases with rotation rate. However, in cases with clouds, the trend is non-monotonic. The temperature difference is higher than the no-cloud case when $\Omgstar = 1/128$, generally decreases with rotation rate until $\Omgstar = 1$, and then increases for higher rotation rates. We suggest that accounting for both clouds and seasonality results in a different trend than that from considering heat transport changes with rotation rate alone.

Considering seasonal variation, we calculated the average equator-to-pole temperature difference for the summer hemisphere in the three months preceding the seasonal transition. In no-cloud simulations, the temperature difference remains relatively constant for $\Omgstar < 1/2$ and increases with rotation rate for $\Omgstar > 1/2$. As depicted in the middle panel of Figure~\ref{fig:Seasonal_Transition}, a higher rotation rate, which corresponds to a lower heat transport efficiency, is associated with a larger phase lag. In our simulations, the temperature difference in no-cloud simulations is significantly higher than in simulations with clouds when $\Omgstar \gtrsim 1/8$. With a higher temperature difference, heat transport efficiency is lower, and consequently, a larger phase lag for no-clouds cases compared to cases with clouds is expected.

However, it is worth noting that for simulations with clouds, the temperature difference remains consistently low ($<$20K) for all rotation rates and decreases with rotation rates for $\Omega^* < 1$, but increases with rotation rates beyond that. Our analysis suggests that heat transport efficiency alone cannot account for the seasonal transition. The radiative timescale also plays a significant role in controlling the seasonal transition \citep{guendelman2022-Key,tan2022-Weak,ohno2019_atmospheres}. Moreover, the radiative feedback may vary with rotation rate in our more realistic radiative and cloud schemes. Thus, further theoretical research that incorporates the evolving radiative feedback with rotation rate is imperative to comprehensively understand the seasonal transition.

\begin{figure*}[htb]
   \centering
   \includegraphics[width=1.0\textwidth]{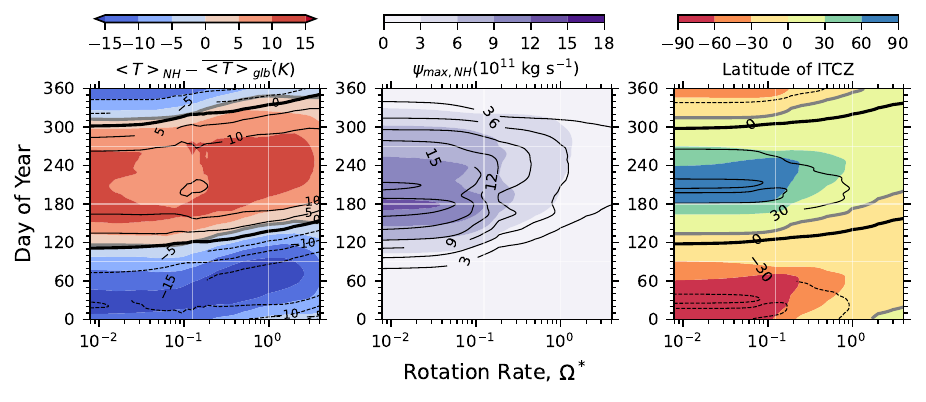}
   \caption{Left: seasonal variation of the temperature, which is defined by the difference between the Northern Hemisphere (NH) averaged surface temperature and the global annual mean surface temperature. Middle: the maximum streamfunction in NH. Right: The latitude of the maximum zonal mean precipitation (ITCZ). The filled shaded contours represent the results from no-cloud simulations and the values are shown in the colourbars, while the open contours represent the results from cloud simulations and the values are shown in inline labels. For comparison, the contours are plotted at the same levels for no-clouds and clouds simulations. To highlight the difference in seasonal transition, the zero contour lines are bold black (clouds) and gray (no-clouds) for the left and right panels. The contours of streamfunction and ITCZ latitude are smoothed by a Gaussian filter with a standard deviation of $\sigma = 3$ days. The horizontal lines represent day of vernal equinox (day 90), northern summer solstice (day 180), and autumnal equinox (day 270), and day 0 is northern winter solstice. The vertical lines represent rotation rates of $\Omgstar$ = 1/8 and 1.}
   \label{fig:Seasons_Effect-contour}
\end{figure*}

To provide more insight into how the rotation rate affects the seasonal transition, we present additional information in Figure~\ref{fig:Seasons_Effect-contour}. Specifically, from left to right, the panels depict the seasonal variation of temperature in the NH, the maximum streamfunction in the NH, and the latitude of the ITCZ for both no-clouds (filled shaded contours) and clouds (open contours) cases, as a function of rotation rate and day of the year. In the left panel, the contours are noticeably tilted upward as the rotation rate increases, consistent with the increases in phase lag shown in the middle panel of Figure~\ref{fig:Seasonal_Transition}. The general trend is that clouds amplify the seasonal amplitude at slower rotation rates, while at $\Omgstar>1/8$, they suppress it. The tilt of the contours with rotation rate is slightly less in the cases with clouds. A distinctive feature occurs in summer around $\Omgstar=1/8$ for no-clouds cases, indicating that the NH average temperature reaches its maximum in early summer and cools rapidly afterwards. This is also shown in the top panel of Figure~\ref{fig:T_diff1}. This feature is specific to no-clouds simulations, and is not related to the kink of albedo and precipitation observed in cloudy cases (Figures~\ref{fig:albedo_seasonal_variation}; \ref{fig:lat-rotation}--\ref{fig:Seasonal_Transition}). The contour plots of the zonal mean temperature at latitudes $40^\circ$N and $60^\circ$N also confirm this feature, with the maximum being reached earlier at lower latitudes, coinciding with the time of ITCZ northward migration.

The middle panel of Figure~\ref{fig:Seasons_Effect-contour} shows the maximum mass streamfunction in the NH. For both cases with and without clouds, the maximum values throughout the year do not vary much with rotation when $\Omgstar<1/8$, and decrease with rotation rate when $\Omgstar>1/8$. The presence of clouds overall increases the maximum streamfunction value, but does not substantially alter its variation with rotation.

A similar trend is found for the latitude of the ITCZ (right panel of Figure~\ref{fig:Seasons_Effect-contour}). The seasonal transition occurs earlier in cases with clouds than without clouds, as discussed above (right bottom panel of Figure~\ref{fig:ITCZ} and middle panel of Figure~\ref{fig:Seasonal_Transition}). The two-peak feature of the ITCZ latitude, shown for the cloudy case with $\Omgstar=1/128$ in Figure~\ref{fig:ITCZ}, can be seen at around day 190 and day 210 for all slowly-rotating cases, and disappears when the rotation rate reaches $\Omgstar\sim1/16$. At intermediate rotation rates, clouds appear to cause solstitial excursions of the ITCZ to occur somewhat earlier in the season, though the modest trend of the equatorial crossing with rotation rate is similar in both cases with and without clouds.

Our findings about the impact of clouds on climate, seasonality, and its changes with rotation rate discussed above all agree qualitatively with the results of previous studies \citep{faulk2017-Effects, guendelman2022-Key, jansen2019-Climates}. While the influence of clouds on seasonality can be attributed to their impact on the energy budget to leading order, it is important to note that our analysis is qualitative. Therefore, further investigation is warranted to gain a more detailed and quantitative understanding of the underlying mechanisms, and to validate our hypotheses.

%% file: 4-discussion.tex
\section{Summary and Discussion}\label{sec:discussion}

Using a suite of simulations with an idealised terrestrial aquaplanet GCM, we demonstrate the relationship between rotation rate, cloud distribution and planetary albedo under the influence of seasonal forcing. We show that rotation rate influences the cloud distribution via the large-scale circulation; consequentially, cloud feedbacks impact the seasonality of the circulation itself. Our key findings are summarised as follows:
\begin{enumerate}
    \item The seasonal cycle leads to a dramatic shift in cloud distribution with rotation rate. At slow rotation rates, cloud area fraction displays a dipolar behaviour near the solstices, as the summer hemisphere is very cloudy whilst the winter hemisphere is relatively cloud-free. At high rotation rates, the clearest distinction between seasons occurs near the equinox rather than the solstice, with the autumnal hemisphere significantly cloudier than the spring hemisphere. The latitudinal distribution of clouds at high rotation rate is more complex with the development of separate equatorial and mid-latitude regions of cloudiness.
    
    \item Cloud distribution is well explained by the large-scale circulation, with the cloudiest regions associated with those undergoing ascent within the atmosphere and having a small (or negative) gradient in equivalent potential temperature. Broadly speaking, low-level clouds occur with a more global distribution, more strongly influenced by the boundary layer.
    
    \item The seasonal cloud distribution greatly impacts the planetary albedo as a function of rotation rate, with substantial inter-seasonal variability. Planetary albedo displays a non-monotonic behaviour that peaks at around $\Omgstar\approx 1/2$, decreasing towards slower rotations as the ITCZ and peak cloud distribution move poleward, and likewise towards faster rotation rates as eddy length scale decreases, reducing the scale of extratropical cyclonic systems that contribute to cloud formation. The local minimum in albedo at $\Omgstar\approx 1/8$ is likely linked to the transition point between a circulation that is mean-flow dominated at slow rotation rates to eddy dominated at high rotation rates.

    \item The introduction of clouds leads to a decrease in annual global mean precipitation compared to cloud-free simulations at all rotation rates, related to their impact on the energy budget. Without clouds, precipitation exhibits a monotonic increase with increasing rotation rates. Once clouds are introduced, a peak in precipitation develops around $\Omgstar\approx 1/8$. Precipitation remains relatively constant for rotation rates below $\Omgstar<1/8$ and increases with rotation rates above $\Omgstar>1/4$.

    \item At slow rotation rates ($\Omgstar<1/8$) the inner edge of the winter Hadley cell and the ITCZ are found near \ang{90} latitude, with little distinction between cloudy and cloud-free cases. At higher rotation rates, a significant difference between the two cases develops, which changes the sign at the fastest rotation rates.

    \item Increasing rotation rate increases the magnitude of the seasonal phase lag in both cloudy and cloud-free simulations. The addition of clouds reduces the seasonal phase lag.

\end{enumerate}

Our results cover a large range of rotation rates, from slow rotators such as Titan to fast rotators such as Earth. This covers a similar parameter space to the simulation grid performed by \citet{kaspi2015-ATMOSPHERIC}, but we do not include simulations equivalent to their fastest ($\Omgstar>4$). This is justified on practical and scientific bases: \citet{chemke2015-Poleward} demonstrated that for fast rotating Earth-like systems, the eddy-driven jets migrate latitudinally even in the absence of seasonal forcing, an effect that does not occur at slower rotation rates. This would be incompatible with any analysis of seasonal averages, particularly with the link demonstrated between large-scale circulation and cloud distribution---we note that our fastest rotating models $\Omgstar \gtrsim 3$ may be affected by this. Running models of higher rotation rates also necessitates the use of higher spectral resolutions and lower model timesteps, imposing a significant penalty on model run-time. At the other end, \citet{yang2014-STRONG} found that slowly rotating planets exhibit strong convergence and convection in the substellar region, leading to the formation of extensive optically thick clouds and a significant increase in planetary albedo. Their results show an increase in planetary albedo with decreasing rotation rates when $\Omgstar<1/8$. However, in our simulations, we did not investigate the tidally-locked atmospheric dynamical regime, even at extremely low rotation rates such as $\Omgstar=1/128$. Our simulations did not exhibit a day-night contrast but rather revealed zonal banded patterns in the daily mean (24-hour mean) climatology.

Regarding how seasonality varies with rotation rate, \cite{faulk2017-Effects} used a more idealised moist GCM, also based on the GFDL spectral dynamical core, to investigate the dependence of ITCZ migration on rotation rate. Their model did not include the effects of clouds and used an idealised gray radiative transfer scheme. As discussed in Section \ref{sec:results-seaonality}, our results generally agree with theirs, except for the maximum latitude of the ITCZ and Hadley cell extent for cases with $\Omgstar<1/8$. \cite{faulk2017-Effects} showed that the ITCZ remains approximately at $\sim\,$\ang{60} while the Hadley cell extent remains at \ang{75}--\ang{80}, while, in our simulations, both the ITCZ and Hadley cell migrate to \ang{90} in cases with and without clouds (Figure \ref{fig:ITCZ} \& \ref{fig:lat-rotation}). We speculate that the difference in the radiation scheme, which leads to insolation and longwave differences, may help determine the location of the ITCZ, which suggests important model dependence. Further, our use of a low mixed-layer depth for the slab ocean allowed the ITCZ to migrate farther poleward in our simulations. Separately, \cite{guendelman2022-Key} and \cite{tan2022-Weak} investigated how the phase lag of seasonal transition varies with rotation rate and concluded that, with a slower rotation rate, the increase in meridional heat transport would decrease the phase lag, which is also seen in our simulations (Figure \ref{fig:Seasonal_Transition} \& \ref{fig:Seasons_Effect-contour}).

We compare our results of simulations with clouds to the results of \citet{komacek2019-Atmospheric}. Komacek \& Abbot used a GCM of higher complexity \citep[ExoCAM;][]{wolf2022-ExoCAM} that includes a correlated-$k$ radiation scheme, allows sea ice to form, and uses sub-grid parameterisations for clouds from \cite{rasch1998-Comparison}. However, their simulations assumed zero obliquity and no seasonal effects. Therefore, the absolute values of streamfunction and jet speed in their results are different from ours. To account for this, we normalise the values to Earth's rotation rate, and we find similar trends: the maximum streamfunction decreases with rotation rate, and the maximum jet speed peaks at around $\Omgstar=1/8$. Our results regarding the latitude of the edge of the Hadley cell and the jet also overlap with those of \citet{komacek2019-Atmospheric}. However, our results on how cloud fraction changes with rotation rate do not agree. In ExoCAM simulations, cloud fraction overall decreases with rotation rate, while our Isca simulations show an increasing trend. There is no clear albedo kink occurring around $\Omgstar=1/8$ in \citet{komacek2019-Atmospheric}, possibly due to their sparse data points on rotation rate but also likely as a consequence of the aseasonality and model configuration differences. The intrinsic difference in cloud parameterisation between the two models, in particular, warrants further research. 

It is likely that there may be differences between our and others' results simply through the use of a particular cloud scheme; \citet{komacek2019-Atmospheric} noted that their results may differ from GCMs using alternative cloud parameterisations. Each cloud scheme is often tuned to reasonably represent the distribution of clouds on present-day Earth. We acknowledge that our simulations present a novel use of the SimCloud scheme, differing primarily from the work by \citet{liu2021-SimCloud} in using an aquaplanet setup and altering the rotation rate. To our understanding, however, none of our changes should invalidate the simple cloud parameterisation itself. Indeed, our study aims at a fundamental understanding of the distribution of clouds in terms of key planetary parameters enabled by a more idealised representation without the complications introduced by effects, such as land-sea heat capacity gradients and orographic uplift, that influence cloud formation.

As well as helping to elucidate the impact of clouds in terrestrial planetary climate systems, our results may be of use to the observation and characterisation of exoplanets. Proposed missions such as the Habitable Worlds Observatory (HWO) aim to study Earth-like exoplanets in reflected light via direct imaging methods. If an Earth-like planet displays an inhomogenous distribution of reflected light (albedo) across the planetary disk, it may be indicative of a cloud-bearing atmosphere \citep{cowan2009-ALIEN, lustig-yaeger2018-Detecting}. With sufficiently long monitoring, it may be possible to infer the rotation rate of the planet by studying changes in the albedo distribution, whose distribution and seasonality we have shown to be related to rotation rate. Regardless, there is ample scope for relating this theoretical work to future observational campaigns.